\documentclass{article}

     \PassOptionsToPackage{numbers, compress}{natbib}

\usepackage[preprint]{neurips_2024}

\usepackage[utf8]{inputenc} 
\usepackage[T1]{fontenc}    
\usepackage{hyperref}       
\usepackage{url}            
\usepackage{booktabs}       
\usepackage{amsfonts}       
\usepackage{nicefrac}       
\usepackage{microtype}      
\usepackage{xcolor}         
\usepackage{amsmath}        
\usepackage{subcaption}
\usepackage{graphicx}
\usepackage{float}
\usepackage{algorithm}
\usepackage{algpseudocode}

\title{Exploring Re-inforcement Learning via Human Feedback under User Heterogeneity}

\author{
  Sarvesh Shashidhar${^1}$, Abhishek Mishra${^1}$, Madhav Kotecha${^2}$ \\
  ${^1}$Centre for Machine Intelligence and Data Science, IIT Bombay\\
  ${^2}$Department of Computer Science and Engineering, IIT Bombay\\
  \texttt{\{24m2152, 24m2156, 24m0833\}@iitb.ac.in}
}

\begin{document}
\maketitle

\begin{abstract}
    Re-inforcement learning from human feedback (RLHF) \cite{rlhf} has been effective in the task of AI alignment. However, one of the key assumptions of RLHF is that the annotators (referred to as \textbf{workers} from here on out) have a \textbf{homogeneous} response space. This assumption is not true in most practical settings and there have been studies done in the past to challenge this notion (\cite{pref9}, \cite{stiennon2020learning}). This work has been inspired by such studies and explores one of the ways to deal with heterogeneity in worker preferences - by \textbf{clustering} workers with similar preferences and personalising reward models for each cluster. This work provides an algorithm that encourages simultaneous learning of reward models and worker embeddings. This algorithm is then empirically tested against the \href{https://huggingface.co/datasets/AbhishekBot/Summarize_Final_Worker_id}{Reddit TL;DR} dataset with unique worker IDs. We have shown that clustering users into different groups based on their preferences and created personalised reward models improves win-rate of the said models. Along with results and visualisations, this work aims to act as a stepping stone to more complicated models and gives a list of possible future extensions.
\end{abstract}

\section{Introduction}
As AI models become more powerful and prominent, aligning them with human preferences has become equally important. \textbf{\textit{Re-inforcement Learning from Human Feedback (RLHF)}} \cite{rlhf} has been a leading approach for AI alignment and has further inspired other algorithms like Direct Preference optimisation (DPO) \cite{dpo}. As the name suggests, RLHF uses human feedback to create \textbf{reward models} that try to maximise the difference in scores between the preferred and rejected responses. Once the reward models have been modelleded (usually as a \textbf{BTL} model \cite{btl}), the AI model's \textbf{\textit{policy}} can now be optimised to match the reward model formulation. \\

The Vanilla RLHF algorithm is quite effective in AI alignment, but at the same time makes a strong assumption - there is \textbf{homogeneity} in human preferences. This means that the RLHF paradigm assumes that all humans shall have the same preference when it comes to a prompt and response pairs. This is however not practically true as human preferences are highly subjective and are shaped by a lot of underlying factors. This large variety of preferences can be seen in all fields (\cite{pref3}, \cite{pref4}, \cite{pref7}, \cite{pref8}) and these \textbf{heterogeneous} preferences should be accounted for during AI alignment. \\

Recent work in this field (\cite{pref9}, \cite{stiennon2020learning}) has helped understand some ways to tackle this challenge. The work done by Park et. al. \cite{pref9} is one such prominent study where the authors proposed ways to create personalised reward models for different users. Inspired by their work, in this study, we shall be trying to handle heterogeneity in user preferences using \textbf{clustering}. The central idea would be to cluster users based on learnt embeddings and then create separate reward models for each of the clusters. This shall help us get a better insight into the origins of heterogeneity and how it can be handled in a non-computationally intensive manner. \\

This study will have some theoretical and algorithmic formulation for the task at hand and these formulations will be backed by empirical experiments. We first start by giving a brief insight into other domains where heterogeneity in human preferences have been observed and then proceed swiftly into some mathematical background for the problem at hand. We shall then explain the methodology and algorithm required and then test the algorithm using empirical results.

\section{Related Work}
In the case of Large Language Models, multiple preference optimisation techniques have been proposed over the years. The first and one of the most trivial ways to perform Preference optimisation was developed by \cite{rlhf}, which used \textbf{Reinforcement Learning using Human Feedback (RLHF)} to optimise preferences. This involved first learning a reward model that would maximise the difference between the scores for the preferred and rejected responses. Post this, the algorithm proceeds to learn a new model policy which is optimal for the learned reward model. \\

Although RLHF, as proposed by \cite{rlhf}, enjoyed initial success, there was a strong underlying assumption in its formulation - \textbf{``human preferences are relatively homogeneous, and can be encoded by a single reward mode''}. This assumption is not true in most cases and causes a serious loss in generality. Human preferences are, by nature, heterogeneous because of the differences in nature, background, upbringing and other factors that shape people's world view and dictate what they prefer. \\

The heterogeneity in human preferences has been the pivotal focus in many studies across domains for decades now. One of the more influential works is the one done by \cite{pref1}, which introduced mixed logit models, a statistical technique that could be used to capture random taste variation across individuals. Building upon this concept, \cite{pref2} proposed an alternative to mixed logit for analysing discrete choices, providing another robust method to identify segments of users with distinct preference structures. These works provided statistical solutions to the heterogeneity problem in human preferences. In addition to this, heterogeneity has also been observed in the field of medicine as explored by \cite{pref3} and \cite{pref4}. These studies provided an understanding of how different patient preferences affected the willingness to participate in studies, treatment and pay for medical intervention. One of the most explored areas for user preferences has been the study of online user preferences. As per \cite{pref5}, there exists inherent variability in how users interact with computing systems and how this variability should be reflected in the design of online websites and resources. Additionally, the work done by \cite{pref6} also analysed online resources and indicated how users with different preferences responded to tasks as they increased in complexity. \\

In recent years, however, most of the studies regarding human preferences have revolved around Artificial Intelligence and Human-Computer Interactions. The survey done by \cite{pref7} shed some light on how individual differences impact the acceptance of AI, confirming that user responses to AI technologies are not uniform and are shaped by various personal characteristics. Additionally, in the field of recommendation systems, work done by \cite{pref8} proposed a user modelling approach that captured fine-grained relatedness between user behaviours using personalised heterogeneous graphs, acknowledging and leveraging the inherent heterogeneity in user interests for improved news recommendation. \\

The study if interest for our work would be the frameworks proposed by \cite{pref9}, where they explored RLHF from Heterogeneous Feedback. Not only did they focus on addressing the issues due to the inherent heterogeneity in human preferences, but also their potential strategic behaviour in providing feedback. Specifically, they propose two frameworks to address heterogeneous human feedback in principled ways: personalisation-based one anda  preference-aggregation-based one. In this work, we shall be focusing on the former method, which shall use ``clustering'' to group users of similar preferences in the embedding space. Then, the model shall proceed to use these similarities to imbibe some homogeneity in this heterogeneous space.

\section{Problem Background}
Preference optimisation algorithms take in preference dataset $\mathcal{D} \hspace{0.1cm} = \hspace{0.1cm} {\left\{ {s_i}, {a_{w}^{i}}, {a_{l}^{i}} \right\}}_{i = 1}^{n}$ where ${s_i} \sim \rho$ is a prompt sampled from the space of possible prompts and ${a_{w}^{i}}, {a_{l}^{i}}$ denote the preferred and rejected responses respectively. Assuming a BTL model as mentioned in \cite{btl} and a latent reward model ${r ^ \ast} \hspace{0.1cm} \colon \hspace{0.1cm} \left( s, a \right) \rightarrow \mathbb{R}$, RLHF (\cite{rlhf}) aims to perform the following 2 steps - 
\begin{enumerate}
    \item It learns an optimal reward model $r ^ \ast$ such that the margin between the scores for $a_w$ and $a_l$ is maximised across prompts and responses.
    \item Then, it uses $r ^ \ast$ to find an optimal policy $\pi ^ \ast$ that maximises the probability of getting $a_{w}^{i}$ for prompt $s^i$ across all $i$
\end{enumerate}

This formulation for RLHF results in the Eq. \ref{eq:3.1}, which gives the optimisation problem for getting the optimal policy - 
\begin{equation}\label{eq:3.1}
    {\pi ^ \ast} \hspace{0.1cm} = \hspace{0.1cm} \arg \max_\pi \hspace{0.1cm} J(\pi) \hspace{0.1cm} = \hspace{0.1cm} \arg \max_\pi \hspace{0.1cm} \mathbb{E}_{s \sim \rho, \hspace{0.05cm} a \sim \pi(. \hspace{0.05cm} \vert \hspace{0.05cm} s)} \hspace{0.1cm} \left[ {r ^ \ast}(s, a) \hspace{0.1cm} - \hspace{0.1cm} \beta \log \left( \frac{\pi(a | s)}{{\pi ^ \text{SFT}} (a | s)} \right)\right]
\end{equation}

This follows from the usage of the BTL model (\cite{btl}) where the probability of choosing the preferred response over the rejected response for a prompt $s$ is given in Eq. \ref{eq:3.2} 
\begin{equation}\label{eq:3.2}
    \mathbb{P} \left( {a_w} \succ {a_l} \hspace{0.1cm} \vert \hspace{0.1cm} s \right) \hspace{0.1cm} = \hspace{0.1cm} \sigma \left[ {r ^ \ast}(s, {a_w}) \hspace{0.1cm} - \hspace{0.1cm} {r ^ \ast}(s, {a_l}) \right]
\end{equation}

However, the above model assumes a homogeneous crowd where each user(worker) has the same preference between any 2 responses for a given prompt. We shall now try to model a similar framework for heterogeneous crowds as well.

\section{Methodology}
The formulation as proposed in Eq. \ref{eq:3.1} is for a homogeneous system of users. However, there is inherent heterogeneity in user preferences (in a practical setting) and there is need to make ``personalised'' reward models. Creating personalised reward models for each user would be impractical while creating a single reward model for all is not optimal either. Therefore, we explore the solution that is a trade-off of both - we resort to \textbf{clustering} \\

Consider a case with $N$ humans/annotators (who shall be referred to as \textbf{workers} here on out) that are divided into $K$ clusters. Each user $i$ has a dataset  $\mathcal{D}_i \hspace{0.1cm} = \hspace{0.1cm} \left\{ \left({o_i^{(j)}} , {\tau_{i,0}^{(j)}} , {\tau_{i, 1}^{(j)}}\right)_{j \in [{N_p}]} \right\}$, where ${o_i^{(j)}}$ is the $j^{\mathrm{th}}$ prompt for the $i ^ {\mathrm{th}}$ user and ${\tau_{i,0}^{(j)}} , {\tau_{i, 1}^{(j)}}$ represent 2 responses to the prompt of user $i$. We shall be assuming that ${\tau_{i,0}^{(j)}}$ and ${\tau_{i,1}^{(j)}}$ are the preferred and rejected responses respectively. Our task is to learn $\theta$ parameters of the model to maximise the margin between the \textbf{log-probabilities} of the preferred and responses while also learning a mapping function $f \hspace{0.1cm} \colon \hspace{0.1cm} [N] \rightarrow [K]$, which dictates the number of clusters optimal for the given $N$ users. Algorithm \ref{alg:1} (as proposed in \cite{pref9}) describes the process we shall be using to achieve these 2 objectives.

\begin{algorithm}[h]
    \caption{personalised RLHF via Clustering}
    \label{alg:1}
    \begin{algorithmic}[1]
        \Require Dataset $\mathcal{D} = \bigcup_{i \in [N]} \hat{\mathcal{D}}_i$ where $\hat{\mathcal{D}}_i = \{(o_i^{(j)}, \tau_{i,0}^{(j)}, \tau_{i,1}^{(j)})\}_{j \in [N_p]}$ is the preference dataset for the $i$th individual.
        
        \State Learn $\theta_{(k)}$ and the clustering map $f: [N] \to [K]$ by:
        \vspace{1ex}
        \Statex
        \begin{align}
            \hat{\theta}_{(k)} &\leftarrow \operatorname{argmax}_{\Vert\theta_{(k)}\Vert_2 \leq B \text{ for all } k \in [K]} \sum_{i \in [N]} \sum_{j \in [N_p]} \log P_{\tilde{\omega}, \theta_{(k)}}(o_i^{(j)} | \tau_{i,0}^{(j)}, \tau_{i,1}^{(j)}) \label{eq:3} \\
            \hat{f}(i) &\leftarrow \operatorname{argmax}_{k \in [K]} \sum_{j \in [N_p]} \log P_{\tilde{\omega}, \theta_{(k)}}(o_i^{(j)} | \tau_{i,0}^{(j)}, \tau_{i,1}^{(j)}) \quad \text{for all } i \in [N]\label{eq:4}
        \end{align}
        \vspace{1ex}
        \For{each $k \in [K]$}
        \State $\hat{\pi}_{(k)} \leftarrow \operatorname{argmax}_{\pi \in \Pi} \left( J(\pi; r_{\tilde{\omega}, \hat{\theta}_{(k)}}) - \mathbb{E}_{\tau \sim \mu_1}[\log \hat{\pi}_{(k)}(\tau)] \right).$
        \EndFor
        
        \Ensure $(\{(\hat{\pi}_{(k)})\}_{k \in [K]}, \{(\hat{\theta}_{(k)})\}_{k \in [K]}, \hat{f})$.
        
    \end{algorithmic}
\end{algorithm}

Algorithm \ref{alg:1} is able to learn the optimal parameters $\hat{\theta}_{(k)}$ and the optimal mapping function $\hat{f}(i)$ to get the optimal number of clusters needed. This has been inspired by the work done by \cite{pref9} where the authors claim that the clustering of users and creating reward models for each cluster will perform better than the Naive RLFH implementation (as proposed by \cite{rlhf}). This claim shall be backed by empirical verification in the next section.

\section{Experimental Setup and Results}
To verify the model proposed in Algorithm \ref{alg:1}, we conduct an empirical evaluation \footnote{The code and other resources can be found \href{https://github.com/AbhishekMishra50/HCAI/tree/main}{here}} of the same on a text summarisation task, using the \textbf{Reddit TL;DR human feedback} dataset(\cite{stiennon2020learning}). The dataset details have been mentioned in \ref{tab:1}.

\begin{table}[h]
    \centering
    \begin{tabular}{|c|c|}
        \hline
        \textbf{Training examples} & 92.9k \\
        \hline
        \textbf{Testing examples} & 86.1k \\
        \hline
        \textbf{Training Workers} & 53 \\
        \hline
        \textbf{Testing Workers} & 63 \\
        \hline
        \textbf{Training Examples  (Filtered)} & 78.5k \\
        \hline
        \textbf{Testing Examples (Filtered)} & 71.5k \\
        \hline
        \textbf{Final Workers} & 40 \\
        \hline
    \end{tabular}
    \caption{Dataset details}
    \label{tab:1}
\end{table}

The data set initially had 53 workers that were used to label training split but there were 63 workers used for labelling the testing split. To ensure consistency, only the rows/prompts with workers common in both splits were filtered out \footnote{The filtered dataset can be found \href{https://huggingface.co/datasets/AbhishekBot/Summarize_Final_Worker_id}{here}}, which resulted in 40 workers in total.

\subsection{Getting User Clusters}
The first step of experimentation was to cluster the workers based on their preferences. To start with a preliminary version of this, we started with a filtered dataset of \textbf{21 workers} and each worker had an embedding of length \textbf{8}. The \verb|GPT-Neo-1.3B| model was then trained for \textbf{5 epochs} over this dataset. Each row of the dataset had a prompt, a worker ID and the preferred and rejected response pairs. \\

Once the AI framework learned the reward models along with the user embeddings, we used the vector of user embeddings to calculate a \textbf{Cosine similarity} between them. This gave us an insight into the similarity of the workers with each other and find workers that are close in the embedding space. To additionally visualize these correlations, we also performed a \textbf{t-distributed Stochastic Neighbor Embedding (t-SNE)} to first reduce the worker embedding dimensions to 2 and then 3 and check how close they are in the embedding space. \\

The same experiments have been repeated for the following cases as well - 
\begin{itemize}
    \item 21 workers with size 16 embeddings
    \item 40 workers with size 16 embeddings
\end{itemize}

The heatmaps and the t-SNE plots have been attach in Appendix \ref{app:a}. Overall, we can observe that there are some correlations between users and we can cluster them into groups using their similarity in the embedding space.

\subsection{Defining personalised Reward Model}
\begin{figure}[h]
    \centering
    \includegraphics[width=\linewidth]{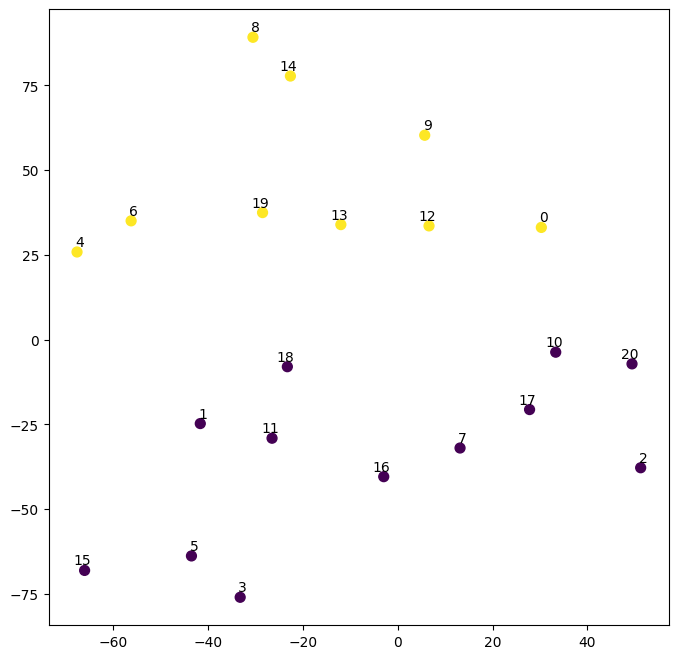}
    \caption{21 users clustered into 2 groups}
    \label{fig:7}
\end{figure}

The personalization of reward model was performed for \textbf{21 workers} that had a size 16 embedding. When all the 21 workers were processed together without any clustering, we were assuming that the model was homogeneous. This is the \textbf{Naive RLHF} model. On the other hand, we also clustered the users into 2 groups based on their embeddings as shown in Fig. \ref{fig:7}. We shall now proceed to create 2 separate reward models for each of the following groups.

The accuracy of the models is measured using the \textbf{win-rate} of the models. The dataset had undergone a \textbf{70-30} split where 70\% of the data was used for training while the 30\% dataset was used for testing. The win-rate of the model is referred to as the percentage of prompt-resoponse pairs where the models is able to correctly identify the preferred and rejected responses. In other words, the higher the win-rate, the more aligned it is to the user preferences. Table \ref{tab:2} shows the Win-rate for the different models.

\begin{table}[h]
    \centering
    \begin{tabular}{|c|c|}
        \hline
        \textbf{Model} & \textbf{Win-Rate (in \%)} \\
        \hline
        Naive RLHF & $\mathbf{52.133\%}$ \\
        \hline
        Group 1 Model (Purple) & $\mathbf{53.221\%}$ \\
        \hline
        Group 2 Model (Yellow) & $\mathbf{52.702\%}$\\
        \hline
    \end{tabular}
    \caption{Win-Rates for the different models}
    \label{tab:2}
\end{table}

\section{Conclusion and Future Work}
Through this study, we were able to implement RLHF as proposed in \cite{rlhf} and examine heterogeneity of the workers in this model. Inspired from the work done in \cite{pref9} and \cite{stiennon2020learning}, we were able to theoretically and empirically examine the presence of heterogeneity in our user base. This enabled us to then create personalised reward models so that reward models are aligned to the preferences of each of the worker groups. We were able to empirically verify that these personalised reward models were able to align better to each of the user groups. \\

This work is preliminary, but is a stepping stone towards understanding human preferences in more detail. As seen in multiple studies (\cite{pref3}, \cite{pref4}, \cite{pref5}, \cite{pref8}), heterogeneity is present in human preferences across domains. Future extension to this work could include optimising reward models for different domains and use cases. To do so however, there is a need to have a comprehensive study about the trade-off between the win-rate gain for the personalised reward models and the computation overhead to create said models. Last but not the least, we could incorporate different metrics to measure the performance of the reward models. Win-rates only measure the accuracy against known preferences, but testing generation capabilities of the models also should be tested against personalization of reward models.

\begin{ack}
We would like to thank Prof. Arpit Agarwal who provided us with the opportunity to pursue this project. This project would not be possible without his guidance and support. We would also like to thank the \textbf{C-MInDS} team, who have supported us by providing the required computation power and technical support to run our highly demanding project code.
\end{ack}

\bibliographystyle{plainnat}
\bibliography{references}

\appendix

\section{Correlation Heatmaps and t-SNE plots}
\label{app:a}
\begin{figure}[h]
    \centering
    \includegraphics[width=0.8\linewidth]{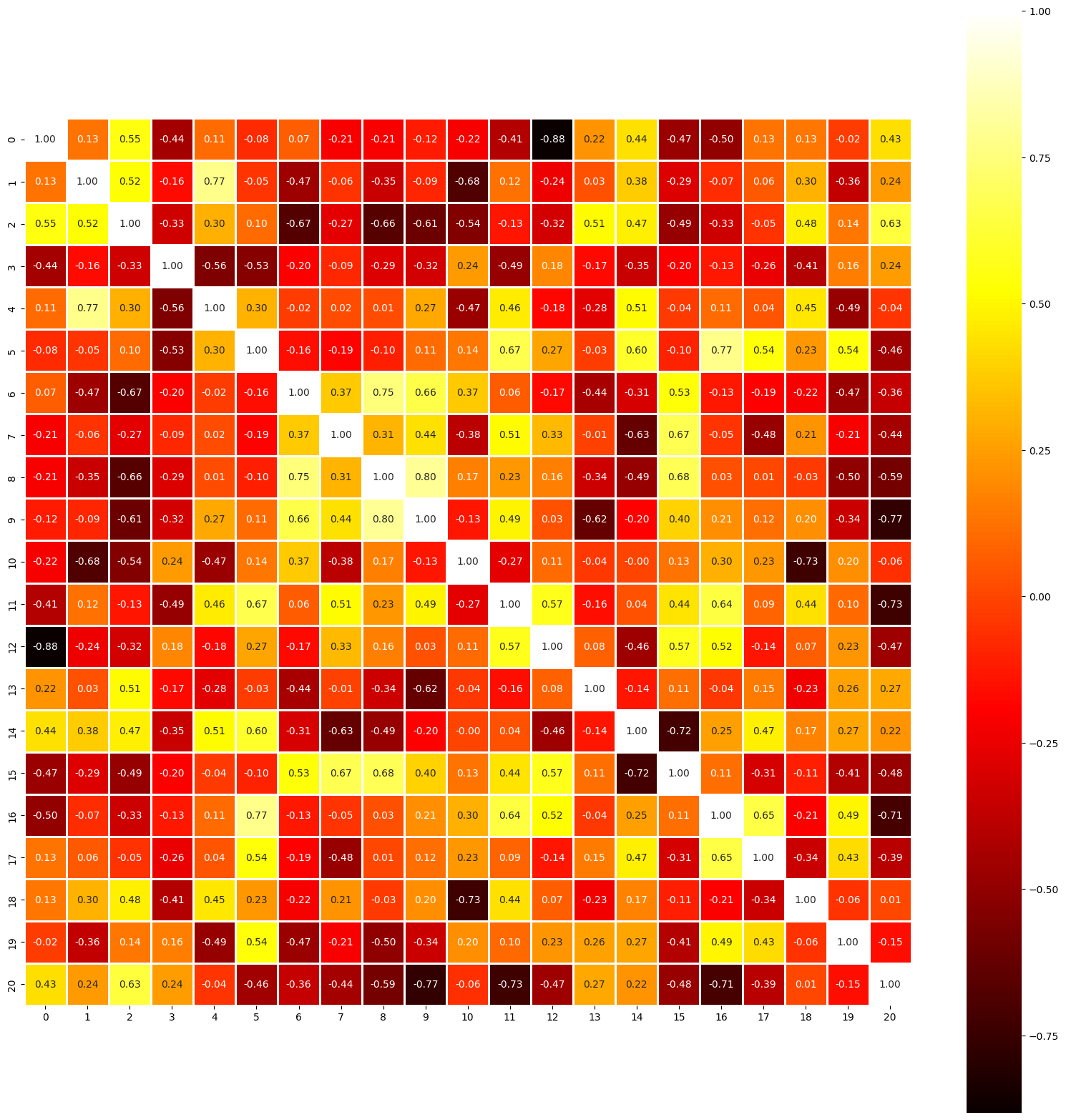}
    \caption{Correlation Heatmap - 21 workers with Size 8 embeddings}
    \label{fig:1}
\end{figure}

Fig. \ref{fig:1} shows the correlation heatmap for the experiment where we took 21 workers and encoded their preferences into vectors of size 8.

We can make the following observations from Fig. \ref{fig:1} - 
\begin{itemize}
    \item Users 8 and 9 are highly positive correlated $(\mathbf{0.80})$
    \item Users 8 and 6 are highly positive correlated $(\mathbf{0.75})$
    \item Users 1 and 4 are highly positive correlated $(\mathbf{0.77})$
    \item Users 0 and 12 are highly negatively correlated $(\mathbf{-0.88})$
    \item User 20 is highly negatively correlated with Users 9$(\mathbf{-0.77})$, 11$(\mathbf{-0.73})$ and 16$(\mathbf{-0.71})$
\end{itemize}

\begin{figure}[H]
    \centering

    \begin{subfigure}[t]{0.47\textwidth}
        \centering
        \includegraphics[width=\linewidth]{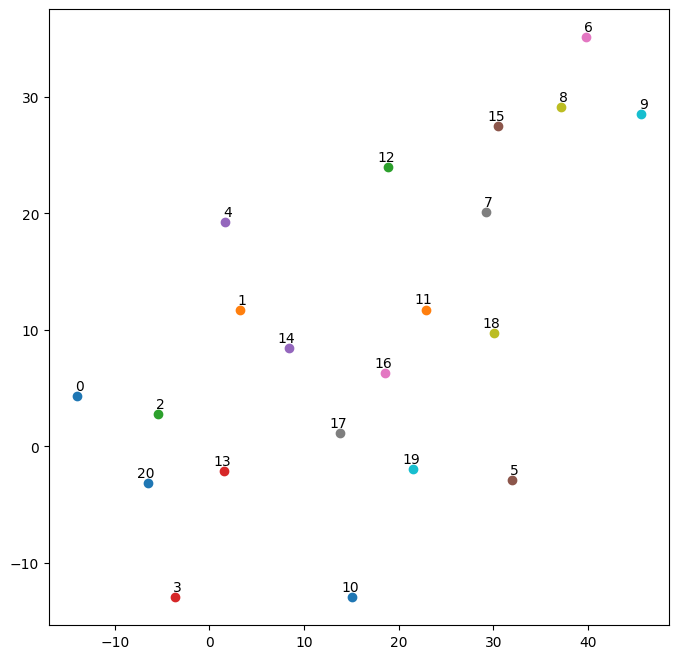}
        \caption{Dimension = 2}
         \label{fig:2_2}
    \end{subfigure}
    \begin{subfigure}[t]{0.47\textwidth}
        \centering
        \includegraphics[width=\linewidth]{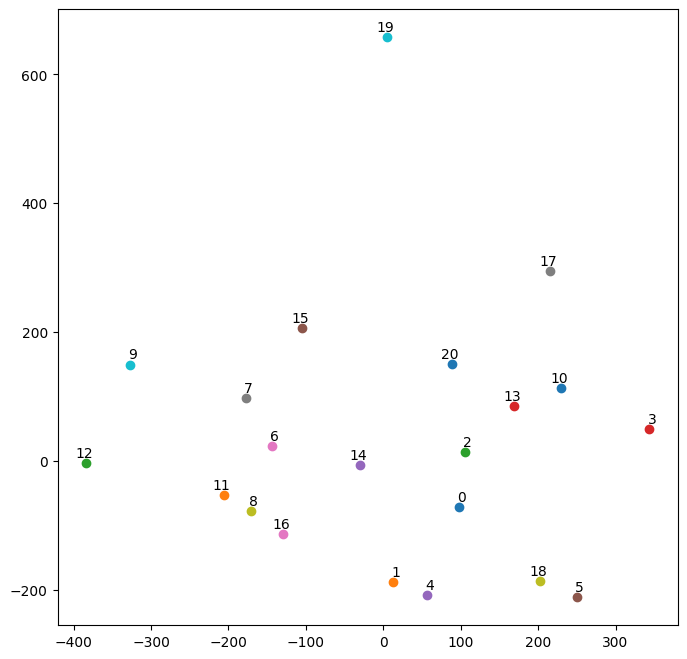}
        \caption{Dimension = 3}
         \label{fig:2_3}
    \end{subfigure}
    \caption{t-SNE Plots for 21 workers and Size 8 embeddings}
    \label{fig:2}
\end{figure}

Fig. \ref{fig:2} shows the t-SNE plots for this experiment by keeping the dimensions as 2 (Fig. \ref{fig:2_2}) and 3 (Fig. \ref{fig:2_3}.

\begin{figure}[h]
    \centering
    \includegraphics[width=0.8\linewidth]{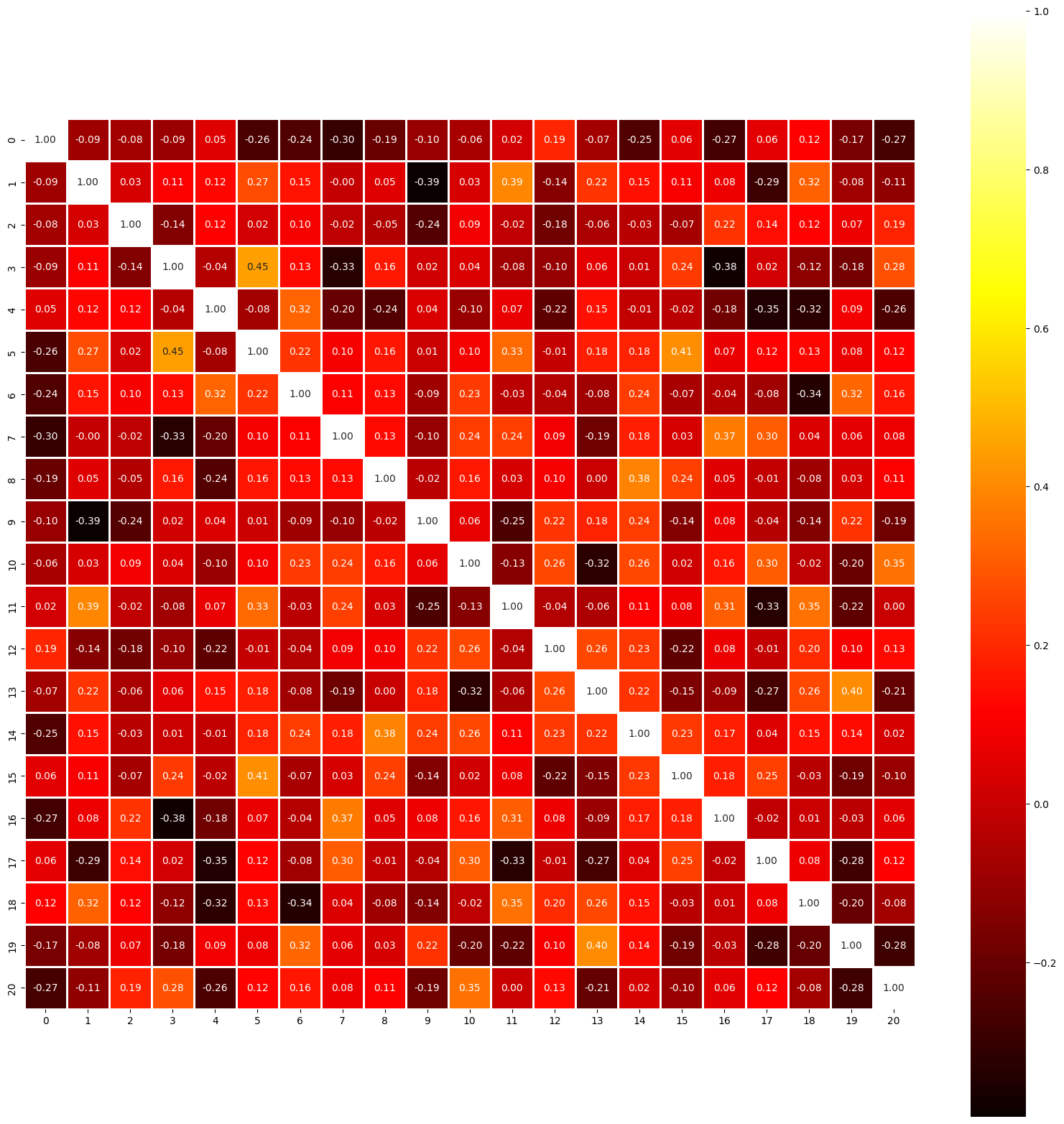}
    \caption{Correlation Heatmap - 21 workers with Size 16 embeddings}
    \label{fig:3}
\end{figure}

Fig. \ref{fig:3} shows the correlation heatmap for the experiment where we took 21 workers and encoded their preferences into vectors of size 16.

We can make the following observations from Fig. \ref{fig:3} - 
\begin{itemize}
    \item Users 3 and 5 are positively correlated $(\mathbf{0.45})$
    \item Users 5 and 15 are positively correlated $(\mathbf{0.41})$
    \item Users 3 and 16 are highly negatively correlated $(\mathbf{-0.38})$
\end{itemize}

\begin{figure}[H]
    \centering

    \begin{subfigure}[t]{0.47\textwidth}
        \centering
        \includegraphics[width=\linewidth]{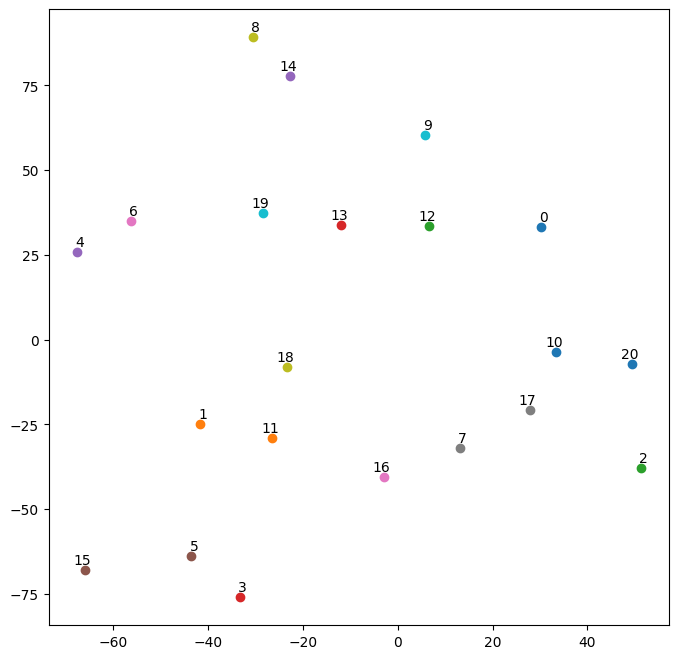}
        \caption{Dimension = 2}
         \label{fig:4_2}
    \end{subfigure}
    \begin{subfigure}[t]{0.47\textwidth}
        \centering
        \includegraphics[width=\linewidth]{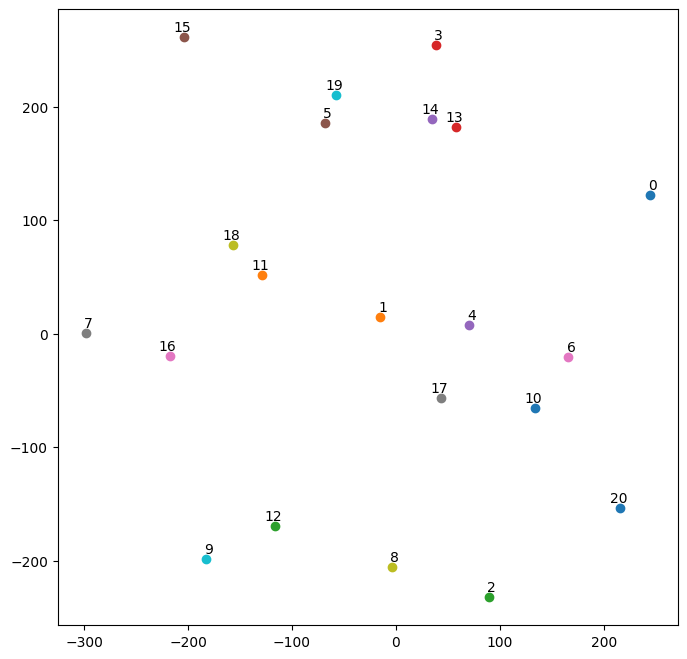}
        \caption{Dimension = 3}
         \label{fig:4_3}
    \end{subfigure}
    \caption{t-SNE Plots for 21 workers and Size 16 embeddings}
    \label{fig:4}
\end{figure}

Fig. \ref{fig:4} shows the t-SNE plots for this experiment by keeping the dimensions as 2 (Fig. \ref{fig:4_2}) and 3 (Fig. \ref{fig:4_3}.

\begin{figure}[h]
    \centering
    \includegraphics[width=0.8\linewidth]{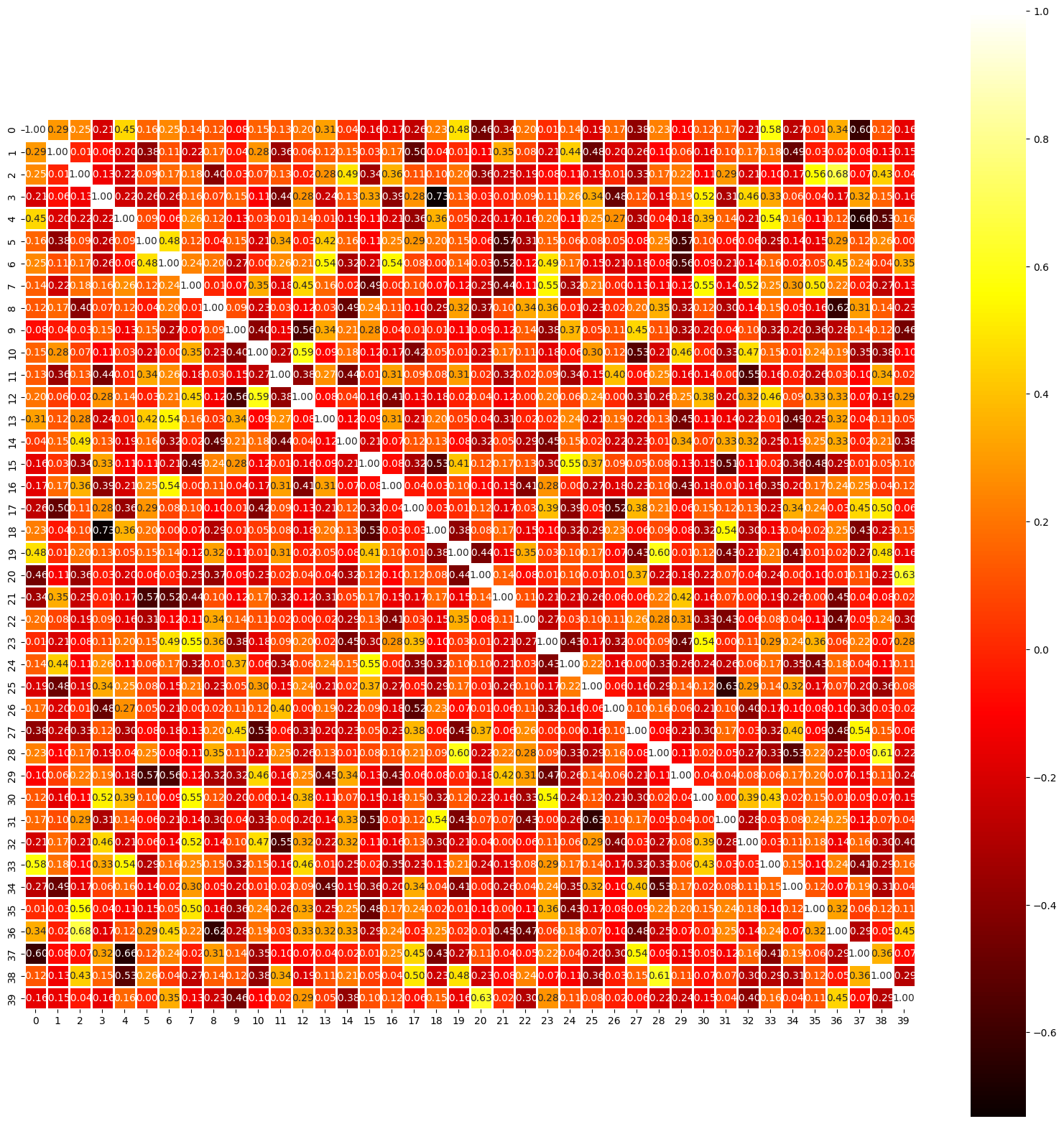}
    \caption{Correlation Heatmap - 40 workers with Size 16 embeddings}
    \label{fig:5}
\end{figure}

Fig. \ref{fig:5} shows the correlation heatmap for the experiment where we took 21 workers and encoded their preferences into vectors of size 8.

We can make the following observations from Fig. \ref{fig:5} - 
\begin{itemize}
    \item Users 2 and 36 are highly positive correlated $(\mathbf{0.68})$
    \item Users 28 and 38 are highly positive correlated $(\mathbf{0.61})$
    \item Users 20 and 39 are highly positive correlated $(\mathbf{0.63})$
    \item Users 3 and 18 are highly negatively correlated $(\mathbf{-0.73})$
\end{itemize}

\begin{figure}[H]
    \centering

    \begin{subfigure}[t]{0.47\textwidth}
        \centering
        \includegraphics[width=\linewidth]{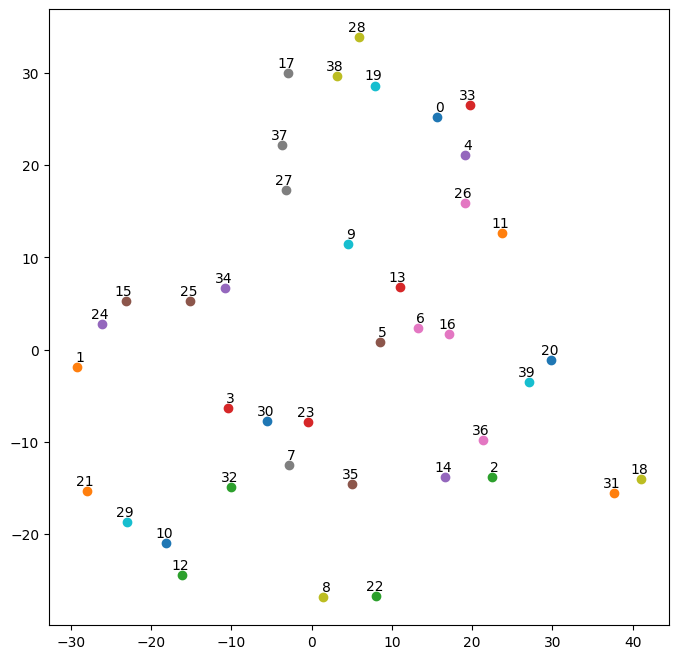}
        \caption{Dimension = 2}
         \label{fig:6_2}
    \end{subfigure}
    \begin{subfigure}[t]{0.47\textwidth}
        \centering
        \includegraphics[width=\linewidth]{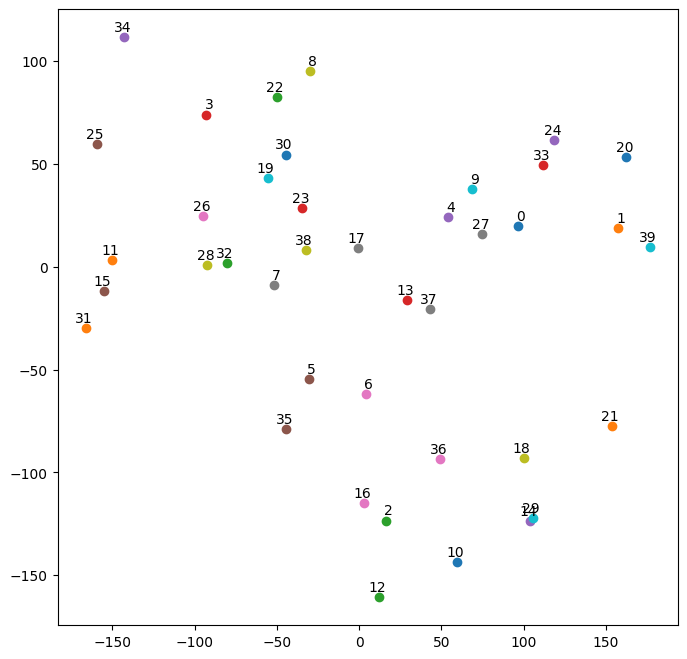}
        \caption{Dimension = 3}
         \label{fig:6_3}
    \end{subfigure}
    \caption{t-SNE Plots for 40 workers and Size 16 embeddings}
    \label{fig:6}
\end{figure}

Fig. \ref{fig:6} shows the t-SNE plots for this experiment by keeping the dimensions as 2 (Fig. \ref{fig:6_2}) and 3 (Fig. \ref{fig:6_3}.

\end{document}